\documentclass{llncs}
\usepackage[numbers,sort&compress,sectionbib]{natbib}
\usepackage{amsmath,graphicx}
\usepackage{amsmath,graphicx}
\usepackage{acro}
\usepackage{tabularx}
\graphicspath{{Figure/}}
\usepackage{color}
\usepackage{multirow}
\usepackage{siunitx}
\usepackage{booktabs} 
\usepackage[font={small}]{caption, subfig}

\usepackage{epstopdf}
\usepackage{ulem}

\newcolumntype{Y}{>{\centering\arraybackslash}X}
\newcommand{\ie}{\textit{i.e.},}
\newcommand{\eg}{\textit{e.g.},}

\newcommand{\etal}{\textit{et al.}}
\newcommand{\Sec}{\S}
\newcommand{\Fig}{Fig.}

\newcommand{\jin}[1]{{#1}} 

\AtBeginDocument{}

\DeclareAcronym{CTV}{
short=CTV,
long= clinical target volume
}

\DeclareAcronym{GTV}{
short=GTV,
long= gross tumor volume
}

\DeclareAcronym{RTCT}{
short=RTCT,
long= radiotherapy computed tomography
}

\DeclareAcronym{RT}{
short=RT,
long= radiotherapy 
}

\DeclareAcronym{PHNN}{
short=PHNN,
long= progressive holistically nested network
}

\DeclareAcronym{LN}{
short=LN,
long= lymph node
}

\DeclareAcronym{OAR}{
short=OAR,
long= organ at risk,
long-plural-form=organs at risk
}

\DeclareAcronym{SDT}{
short=SDM,
long= signed distance transform map
}

\DeclareAcronym{CNN}{
short=CNN,
long= convolutional neural network
}

\DeclareAcronym{CT}{
short=CT,
long= computed tomography
}

\DeclareAcronym{DS}{
short=DS,
long= Dice score
}

\DeclareAcronym{ASD}{
short=ASD,
long= average surface distance
}

\DeclareAcronym{HD}{
short=HD,
long= Hausdorff distance
}

\DeclareAcronym{FCN}{
short=FCN,
long= fully convolutional network
}
\DeclareAcronym{VOI}{
short=VOI,
long= volume of interest
}

\title{Deep Esophageal Clinical Target Volume Delineation using Encoded 3D Spatial Context of Tumors, Lymph Nodes, and Organs At Risk}

\author{Dakai Jin\textsuperscript{1} \and Dazhou Guo\textsuperscript{1} \and Tsung-Ying Ho\textsuperscript{2} \and Adam P. Harrison\textsuperscript{1} \and Jing Xiao\textsuperscript{3} \and Chen-kan Tseng\textsuperscript{2} \and Le Lu\textsuperscript{1} }

\institute{\textsuperscript{1}PAII Inc., Bethesda, MD, USA \\ \textsuperscript{2}Chang Gung Memorial Hospital, Linkou, Taiwan, ROC \\ \textsuperscript{3}Ping An Technology, Shenzhen, China \\
}

\begin{document}
\setlength{\abovecaptionskip}{1ex}
\setlength{\belowcaptionskip}{1ex}
\setlength{\floatsep}{1ex}

\maketitle

\begin{abstract}
\Ac{CTV} delineation from \ac{RTCT} images is used to define the treatment areas containing the \ac{GTV} and/or sub-clinical malignant disease for \ac{RT}. High intra- and inter-user variability makes this a particularly difficult task for esophageal cancer. This motivates automated solutions, which is the aim of our work. Because \ac{CTV} delineation is highly context-dependent---it must encompass the \ac{GTV} and regional \acp{LN} while also avoiding excessive exposure to the \acp{OAR}---we formulate it as a deep contextual appearance-based problem using encoded spatial contexts of these anatomical structures. This allows the deep network to better learn from and emulate the margin- and appearance-based delineation performed by human physicians. Additionally, we develop domain-specific data augmentation to inject robustness to our system. Finally, we show that a simple 3D \ac{PHNN}, which avoids computationally heavy decoding paths while still aggregating features at different levels of context, can outperform more complicated networks.  Cross-validated experiments on a dataset of $135$ esophageal cancer patients demonstrate that our encoded spatial context approach can produce concrete performance improvements, with an average Dice score of $83.9\pm5.4\%$ and an average surface distance of $4.2\pm2.7\,\mathrm{mm}$, representing improvements of $3.8\%$ and $2.4 \mathrm{mm}$, respectively, over the state-of-the-art approach. 


\end{abstract}


\acresetall

\section{Introduction}
\label{sec:intro}
\begin{figure}[t]
\centering
\includegraphics[scale=.25]{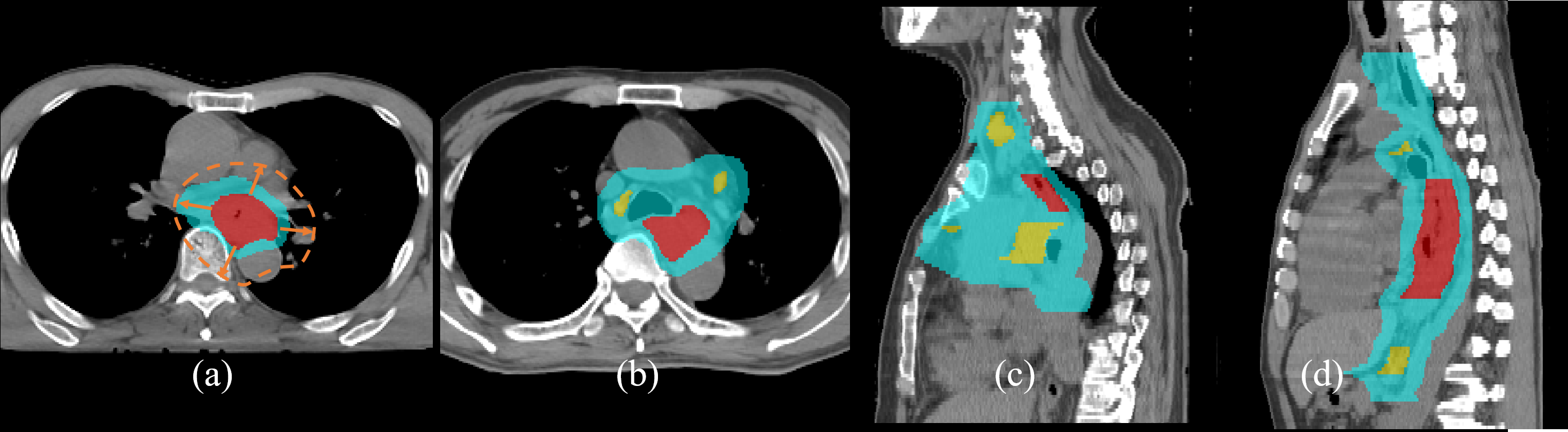}
\caption{Esophageal cancer \acs{CTV} delineation, where red, yellow, and cyan indicate the \acs{GTV}, regional \acsp{LN} and \acs{CTV}, respectively. (a) shows that the \acs{CTV} is not a uniform margin expansion (brown-dotted line) from the \acs{GTV}, while (b)-(d) shows how delineation becomes more complicated when regional \acsp{LN} are present. (c) and (d) also depict wide and long examples of esophageal \acs{CTV}, respectively.}
\label{fig:motivation}
\end{figure}

Esophageal cancer ranks the sixth in global cancer mortality~\cite{bray2018global}. As it is usually diagnosed at rather late stage~\cite{zhang2013epidemiology}, \ac{RT} is a cornerstone of treatment. Delineating the 3D \ac{CTV} on a \ac{RTCT} scan is a key challenge in \ac{RT} planning. As \Fig~\ref{fig:motivation} illustrates, the \ac{CTV} should spatially encompass, with a mixture of predefined and judgment-based margins, primary tumor(s), \ie{} the \ac{GTV}, regional \acp{LN} and sub-clinical disease regions, while simultaneously limiting radiation exposure to \acp{OAR}~\cite{burnet2004defining}. 

Esophageal \ac{CTV} delineation is uniquely challenging because tumors may potentially spread along the entire esophagus and metastasize up to the neck or down to the upper abdomen \acp{LN}. Current clinical protocols rely on manual \ac{CTV} delineation, which is very time and labor consuming and is subject to high inter- and intra-observer variability~\cite{louie2010inter}. This motivates automated approaches to the \ac{CTV} delineation.

Deep \acp{CNN} have achieved notable successes in segmenting semantic objects, such as organs and tumors, in medical imaging~\cite{Cicek2016,jin20173d,harrison2017progressive,jin2018ct, heinrich2019obelisk,jin2019gtv}. However, to the best of our knowledge, no prior work, \ac{CNN}-based or not, has addressed esophageal cancer \ac{CTV} segmentation. Works on \ac{CTV} segmentation of other cancer types mostly operate based on the \ac{RTCT} appearance alone~\cite{men2017automatic, men2018fully}. As shown in Fig.~\ref{fig:motivation}, \ac{CTV} delineation depends on the radiation oncologist's visual judgment of both the appearance \textit{and} the spatial configuration of the \ac{GTV}, \acp{LN}, and \acp{OAR}, suggesting that only considering the \ac{RTCT} makes the problem ill-posed. Supporting this, Cardenas \etal{} recently showed that considering the \ac{GTV} and \ac{LN} binary masks together with the \ac{RTCT} can boost oropharyngeal \ac{CTV} delineation performance~\cite{cardenas2018auto}. However, the \acp{OAR} were not considered in their work. Moreover, binary masks do not explicitly provide distances to the model. Yet \ac{CTV} delineation is highly driven by distance-based margins to other anatomical structures of interest, and it is difficult to see how regular \acp{CNN} could capture these precise distance relationships with binary masks alone. 

Our work fills this gap by introducing a spatial-context encoded deep \ac{CTV} delineation framework. Instead of expecting the \ac{CNN} to learn distance-based margins from the \ac{GTV}, \ac{LN}, and \ac{OAR} binary masks, we provide the \ac{CTV} delineation network with the 3D \acp{SDT}~\cite{Sethian1996Fast} of these structures. Specifically, we include the \acp{SDT} of the \ac{GTV}, \acp{LN}, lung, heart, and spinal canal with the original \ac{RTCT} volume \jin{as inputs to the network}. From a clinical perspective, this allows the \ac{CNN} to emulate the oncologist's manual delineation, which uses the distances of \ac{GTV} and \acp{LN} vs. the \acp{OAR} as a key constraint in determining \ac{CTV} boundaries. To improve robustness, we randomly choose manually and automatically generated \ac{OAR} \acp{SDT} during training, while augmenting the \jin{\ac{GTV} and \acp{LN}} \acp{SDT} with the domain-specific jittering. We adopt a 3D \ac{PHNN}~\cite{harrison2017progressive} to serve as our delineation model, which enjoys the benefits of strong abstraction capacities and multi-scale feature fusion \jin{with a light-weighted} decoding path. We extensively evaluate our approach using a 3-fold cross-validated dataset of $135$ esophageal cancer patients. Since we are the first to tackle automated esophageal cancer \ac{CTV} delineation, we compare against previous \ac{CTV} delineation methods for other cancers~\cite{men2018fully,cardenas2018auto}, using the 3D \ac{PHNN} as the delineation model. When comparing against pure appearance-based~\cite{men2018fully} and binary-mask-based~\cite{cardenas2018auto} solutions, we show that our approach provides improvements of $10\%$ and $3.8\%$ in Dice score, respectively, with analogous improvements in \ac{HD} and \ac{ASD}. Moreover, we also show that \ac{PHNN} is responsible for providing improvements of $1\%$ in Dice score and $0.4mm$ reduction in \ac{ASD} over a 3D U-Net model~\cite{Cicek2016}. 


\section{Methods}
\label{sec:method}
\begin{figure}[ht]
\centering
\includegraphics[width=0.99\textwidth]{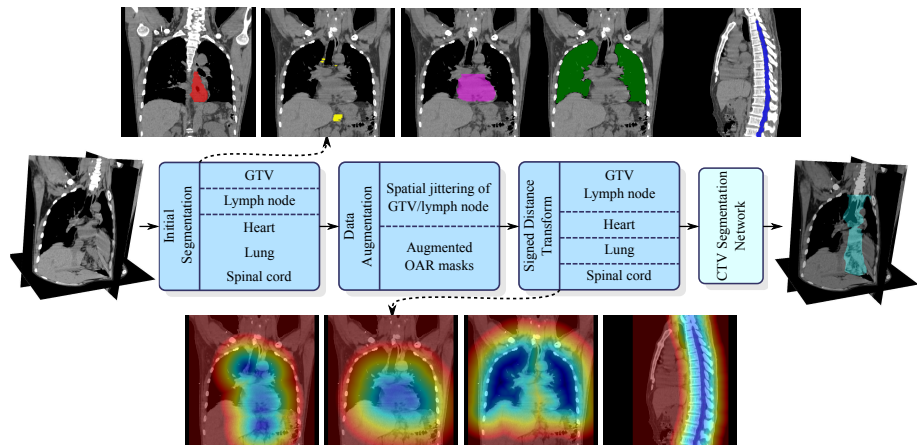}
\caption{Overall workflow of our spatial context encoded \acs{CTV} delineation framework. The top and bottom rows depict different masks and \acp{SDT}, respectively, overlayed on the \ac{RTCT}. From left to right are the \ac{GTV}, \acp{LN}, heart, lung, and spinal canal. The \ac{GTV} and \acp{LN} share a combined \ac{SDT}.}
\label{fig:ctv_workflow}
\end{figure}

\ac{CTV} delineation in \ac{RT} planning is essentially a margin expansion process, starting from observable tumorous regions (\ac{GTV} and regional \acp{LN}) and extending into the neighboring regions by considering the possible tumor spread margins and distances to nearby healthy \acp{OAR}. \Fig~\ref{fig:ctv_workflow} depicts an overview of our method, which consists of four major modularized components: (1) segmentation of prerequisite regions; (2) \ac{SDT} computation; (3) domain-specific data augmentation; and (4) a 3D \ac{PHNN} to execute the \ac{CTV} delineation.

\subsection{Prerequisite Region Segmentation}
\label{sec:pre_seg}
To provide spatial context/distance of the anatomical structures of interest, we must first know their boundaries. We assume that manual segmentations for the esophageal \ac{GTV} and regional \acp{LN} are available. However, we do not make this assumption for the \acp{OAR}. Indeed, missing \ac{OAR} segmentations ($\sim 20\%$) is common in our dataset. For the \acp{OAR}, we consider three major organs: the lung, heart, and spinal canal, since most esophageal \acp{CTV} are closely integrated with these organs. Using the available organ labels, we trained a 2D \ac{PHNN}~\cite{harrison2017progressive} to segment the \acp{OAR}, considering its robust performance in pathological lung segmentation and its computational efficiency. Examples of automatic \ac{OAR} segmentation are illustrated in the first row in \Fig~\ref{fig:ctv_workflow} and  validation Dice score for the lung, heart and spinal canal were $97\%$, $95\%$ and $78\%$, respectively, in our dataset.

\subsection{\acs{SDT} Computation} 
To encode the spatial context with respect to the \ac{GTV}, regional \acp{LN}, and \acp{OAR}, we compute \acfp{SDT} for each.  The \ac{SDT} is generated from a binary image, where the value in each voxel measures the distance to the closest object boundary. Voxels inside and outside the boundary have positive and negative values, respectively. More formally, let  $\mathcal{O}_{i}$ denote a binary mask, where $i\in \{ \text{GTV+LNs, lung, heart, spinal canal} \}$ and let $\Gamma(\cdot)$ be a function that computes boundary voxels of a binary image. The \ac{SDT} value at a voxel $p$ with respect to $\mathcal{O}_i$ is computed as

\begin{align}
\text{SDM}_{\Gamma(\mathcal{O}_i)}(p) = \left \{
\begin{array}{rcl}
\underset{q\in \Gamma(\mathcal{O}_i)}{\min} d(p,q)  & \quad {\text{if} \quad p\notin \mathcal{O}_i}\\
-\underset{q\in \Gamma(\mathcal{O}_i)}{\min} d(p,q)  & \quad {\text{if} \quad p\in \mathcal{O}_i }
\end{array} \right. \mathrm{,}
\end{align}
where $d(p,q)$ is a distance measure from $p$ to $q$. We choose to use Euclidean distance in our work and use Maurer \etal{}'s efficient algorithm~\cite{maurer2003linear} to compute the \acp{SDT}. The bottom row in \Fig~\ref{fig:ctv_workflow} depicts example \acp{SDT} for the combined \ac{GTV} and \acp{LN} and the other 3 \acp{OAR}. Note that we compute \acp{SDT} separately for each of the three \acp{OAR}, meaning we can capture each organ's influence on the \ac{CTV}. Providing the \acp{SDT} of the \ac{GTV}, \acp{LN}, and \acp{OAR} to the deep \ac{CNN} allows it to more easily  \jin{infer} the distance-based margins to these anatomical structures, better emulating the oncologist's \ac{CTV} inference process.

\subsection{Domain-Specific Data Augmentation}
We adopt specialized data augmentations to increase the robustness of the training and harden our network to noise in the prerequisite segmentations. Specifically, two types of data augmentation are carried out. (1) We calculate the \ac{GTV} and \acp{LN} \acp{SDT} from both the manual annotations and also spatially jittered versions of those annotations. We jitter each \jin{\ac{GTV} and \ac{LN} component} by random shift within $4 \times 4 \times 4\, \mathrm{mm}^3$, mimicking that in practice $4\,\mathrm{mm}$ average distance error represents the state-of-the art performance in esophageal \ac{GTV} segmentation~\cite{yousefi2018esophageal,jin2019gtv}. (2) We calculate \acp{SDT} of the \acp{OAR} using both the manual annotations and the automatic segmentations from \Sec\ref{sec:pre_seg}. Combined, these augmentations lead to four possible combinations, which we randomly choose between during every training epoch. This increases model robustness and also allows the system to be effectively deployed in practice \jin{by using \acp{SDT} of the automatically segmented \acp{OAR}} , helping to alleviate the labor involved. 

\subsection{CTV Delineation Network}

To use 3D \acp{CNN} in medical imaging, one has to strike a balance between choosing the appropriate image size covering enough context and the GPU memory.  The symmetric encoder-decoder segmentation networks, \eg{} 3D U-Net~\cite{Cicek2016}, are computationally heavy and memory-consuming since half of its computation is consumed on the decoding path, which may not always be needed for all 3D segmentation tasks. To alleviate the computational/memory burden, we adopt a 3D version of \ac{PHNN}~\cite{harrison2017progressive} as our \ac{CTV} delineation network, which is able to fuse different levels of features using parameter-less deep supervision. We keep the first 4 convolutional blocks and adapt it to 3D as our network structure. As we demonstrate in the experiments, the 3D \ac{PHNN} is not only able to achieve reasonable improvement over the 3D U-Net but requires 3 times less GPU memory.

\section{Experiments and Results}
\label{sec:experiments}

To evaluate the performance of our esophageal \ac{CTV} delineation framework, we collected from $135$ anonymized \acp{RTCT} of esophageal cancer patients undergoing \ac{RT}. Each \ac{RTCT} is accompanied by a \ac{CTV} mask annotated by an experienced oncologist, based on a previously segmented \ac{GTV}, regional \acp{LN}, and \acp{OAR}.  The average \ac{RTCT} size is $512\times512\times250$ voxels with the average resolution of $1.05\times1.05\times2.6$ mm.  

\noindent\textbf{Training data sampling:}
We first resample all the \acs{CT} and \ac{SDT} images to a fixed resolution of $1.0\times1.0\times2.5$ mm, from which we extract $96 \times 96 \times 64$ training \ac{VOI} patches in two manners: (1) To ensure enough \acp{VOI} with positive \ac{CTV} content, we randomly extract \acp{VOI} centered within the \ac{CTV} mask. (2) To obtain sufficient negative examples, we randomly sample $\sim20$ \acp{VOI} from the whole volume.  This results in on average $80$ \acp{VOI} per patient. We further augment the training data by applying random rotations of $\pm10$ degrees in the x-y plane.

\noindent\textbf{Implementation details:} The Adam solver~\cite{kingma2014adam} is used to optimize all segmentation models with a momentum of $0.99$ and a weight decay of $0.005$ for $30$ epochs. We use the Dice loss for training. For testing, we use 3D sliding windows with sub-volumes of $96 \times 96 \times 64$ and strides of $64 \times 64 \times 32$ voxels. The probability maps of sub-volumes are aggregated to obtain the whole volume prediction taking on average \jin{$6-7\,\mathrm{s}$} to process one input volume using a Titan-V GPU. 

\noindent\textbf{Comparison setup and metrics:}
We use 3-fold cross-validation, separated at the patient level, to evaluate performance of our approach and the competitor methods. We compare against setups using only the  \ac{CT} appearance information~\cite{men2017automatic,men2018fully} and setups using the \ac{CT} with binary \ac{GTV}/\ac{LN} masks~\cite{cardenas2018auto}. Finally, we also compare against setups using the \ac{CT} + \ac{GTV}/\ac{LN} \acp{SDT}, which does not consider the \acp{OAR}. We compare these setups using the 3D \ac{PHNN}. For the 3D U-Net~\cite{Cicek2016}, we compared against the setup using the \ac{CT} appearance information.  We evaluate the performance using the metrics of Dice score, \ac{ASD} and \ac{HD}.

\begin{figure}
\centering
\includegraphics[width=0.93\textwidth]{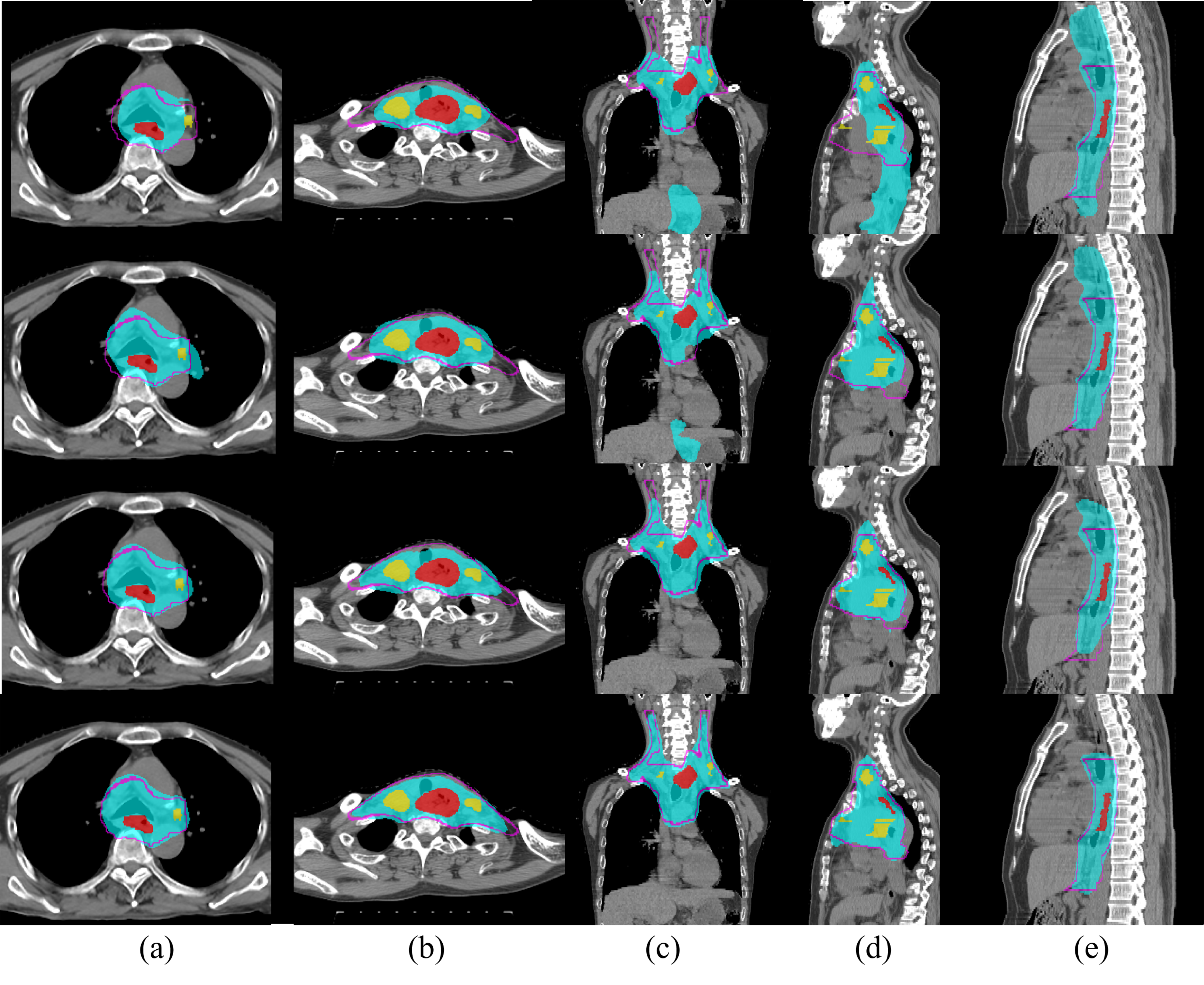}
\caption{Qualitative illustration of esophageal \acs{CTV} delineation using different \ac{PHNN} setups. Red, yellow and cyan represent the \ac{GTV}, \ac{LN} and predicted \ac{CTV} regions, respectively. The purple line indicates the ground truth \ac{CTV} boundary. The $1^{st}$ and $2^{nd}$ rows show examples from setups using pure \ac{RTCT}~\cite{men2018fully} and when adding \ac{GTV}/\ac{LN} binary masks~\cite{cardenas2018auto}, respectively. The $3^{rd}$ and $4^{th}$ row show examples when adding \ac{GTV}/\ac{LN} \acp{SDT} and our proposed \ac{GTV}/\ac{LN}/\ac{OAR} \acp{SDT}, respectively. (a) and (d) demonstrate that the pure \ac{RTCT} setups fail to include the regional \acp{LN}, while (c) to (e) depict severe over-segmentations. While these errors are partially addressed using the \ac{GTV}/\ac{LN} mask setup, it still suffers from inaccurate \ac{CTV} boundaries (a-c) or over coverage of normal regions (d,e). These issues are much better addressed by our proposed method. }
\label{fig:ctv_quality}
\end{figure}

\begin{table}[]
\centering
\caption{Quantitative results for the esophageal cancer \ac{CTV} delineation. The last, starred row represents performance when using automatically generated \ac{OAR} \acp{SDT}. }\label{tbl:results}

\begin{tabular}{l|c|ccc}
Models                      &       Setups                   & Dice           & \ac{HD} (mm)         & ASD (mm)      \\ \hline
\multirow{2}{*}{U-Net} & CT                              & 0.739$\pm$0.126  & 69.5$\pm$42.7 & 10.1$\pm$9.4  \\
                      & CT + GTV/LN/OAR \acsp{SDT}      & 0.829$\pm$0.061   & 36.9$\pm$23.8 & 4.6$\pm$3.0   \\ \hline
\multirow{5}{*}{PHNN} & CT                              & 0.739$\pm$0.117 & 68.5$\pm$43.8 & 10.6$\pm$9.2   \\
                      & CT + GTV/LN masks              & 0.801$\pm$0.075 & 56.3$\pm$35.4 & 6.6$\pm$5.3   \\
                      & CT + GTV/LN \acsp{SDT}                  & 0.816$\pm$0.067  & 44.7$\pm$25.1 & 5.4$\pm$4.1   \\
                      & CT + GTV/LN/OAR \acsp{SDT}             & \textbf{0.839$\pm$0.054} & \textbf{35.4$\pm$23.7} & \textbf{4.2$\pm$2.7}   \\
                      & CT + GTV/LN/OAR \acsp{SDT}*   &     0.823$\pm$0.059 &     43.6$\pm$26.4  & 5.1$\pm$3.3   \\ \hline
\end{tabular}
\end{table}

\begin{figure}
\centering
\includegraphics[width=0.95\textwidth]{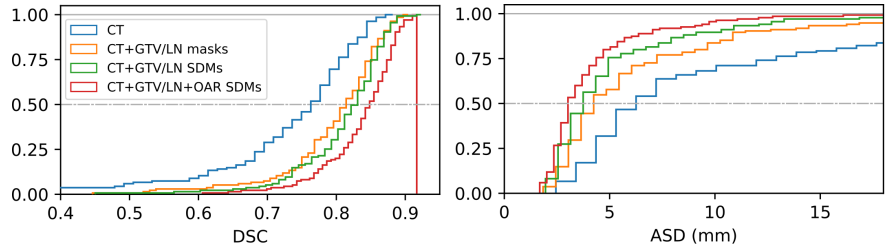}
\caption{Cumulative histograms of the CTV delineation performance under 4 setups using 3D PHNN on cross-validated 135 patients. The left and right depict the Dice score and \ac{ASD} results, respectively. From the results, we observe that $> 77\%$ patients have Dice score $\geq 0.80$, and $> 55\%$ patients have Dice score $\geq 0.85$ by using the proposed method (shown in red). Since there are often large inter-observer variations on CVT delineation tasks, \jin{\ie{} ranging from 0.51 to 0.81 in terms of Jaccard index in cervix cancer~\cite{eminowicz2015variability}}, these findings may indicate that, for a high percentage of the studied patient population, little to no additional manual revision is needed on the automatically delineated CTVs.}
\label{fig:hist} 
\end{figure}

\noindent\textbf{Results:}
Table 1 outlines the quantitative comparisons of the different model setups and choices. As can be seen, methods based on pure \ac{CT} appearance, seen in prior art~\cite{men2017automatic,men2018fully}, exhibits the worst performance. This is  because inferring distance-based margins from appearance alone is too hard of a task for \acp{CNN}. Focusing on the \ac{PHNN} performance, when adding the binary \ac{GTV} and \ac{LN} masks as contextual information~\cite{cardenas2018auto}, the performance increases considerably from $0.739\pm0.117$ to $0.801\pm0.075$ in Dice score. When using the \ac{SDT} encoded spatial context of \ac{GTV}/\ac{LN}, \ac{PHNN} further improves the Dice score and \ac{ASD} by $1.5\%$ and $1.2\,\mathrm{mm}$, respectively, confirming the value of using the distance information for esophageal \ac{CTV} delineation. Finally, when the \ac{OAR} \acp{SDT} are included, \ie{} our proposed framework, \ac{PHNN} achieves the best performance reaching $0.839\pm0.054$ Dice score and $4.2\pm2.7\,\mathrm{mm}$ \ac{ASD}, with a reduction of $9.3\,\mathrm{mm}$ in \ac{HD} as compared to the next best \ac{PHNN} result.  \Fig~\ref{fig:hist} depicts cumulative histograms of the Dice score and \ac{ASD}, visually illustrating the distribution of improvements in the CTV delineation performance.  \Fig~\ref{fig:ctv_quality} shows some qualitative examples illustrating these performance improvements. Interestingly, as the last row of Table 1 shows, when using \acp{SDT} computed from the automatically segmented \acp{OAR} for testing, the performance compares favorably to the best configuration, and outperforms all other configurations. This indicates that our method remains robust to noise within the \ac{OAR} \acp{SDT} and also that our approach is not reliant on manual \ac{OAR} masks for good performance, increasing its practical value.

We also compare the 3D \ac{PHNN} network performance with that of 3D U-Net~\cite{Cicek2016} when using the \ac{CT} appearance based setup and the proposed whole framework. As Table 1 demonstrates, when using the whole pipeline \ac{PHNN} outperforms U-Net by $1\%$ dice score. Although \ac{PHNN} has similar performance against U-Net when using only the \ac{CT} appearance information, the GPU memory consumption is roughly 3 times less than that of the U-Net. These results indicate that for esophageal \ac{CTV} delineation, a \ac{CNN} equipped with strong encoding capacity and a light-weight decoding path can be as good as (or even superior to) a heavier network with a symmetric decoding path.

\section{Conclusion}
\label{sec:conclusion}
We introduced a spatial-context encoded deep esophageal \ac{CTV} delineation framework designed to produce superior margin-based \ac{CTV} boundaries. Our system encodes spatial context by computing the \acp{SDT} of the \ac{GTV}, \acp{LN} and \acp{OAR} and feeds them together with the \ac{RTCT} image into a 3D deep \ac{CNN}. Analogous to clinical practice, this allows the system to consider both appearance and distance-based information for delineation. Additionally, we also developed domain-specific data augmentation and adopted a 3D \ac{PHNN} to further improve robustness. Using extensive three-fold cross-validation, we demonstrated that our spatial-context encoded approach can outperform state-of-the-art \ac{CTV} alternatives by wide margins in Dice score, \ac{HD}, and \ac{ASD}. As we are the first to address automated esophageal \ac{CTV} delineation, our method represents an important step forward for this important problem.

\bibliographystyle{splncs}
\small{\bibliography{refs}}
\end{document}